\documentclass[11pt]{article}

\usepackage[a4paper,margin=1in]{geometry}
\usepackage{amsmath,amssymb,amsthm,mathtools}
\usepackage{booktabs}
\usepackage{microtype}
\usepackage[T1]{fontenc}
\usepackage{lmodern}
\usepackage{graphicx}
\usepackage{flafter}
\usepackage{listings}
\usepackage{xcolor}
\usepackage{hyperref}

\graphicspath{{figures/}}

\hypersetup{colorlinks=true,linkcolor=blue,citecolor=blue,urlcolor=blue}

\newcommand{\trans}{^{\top}}
\newcommand{\Sig}{\Sigma}

\title{The Efficient Frontier from a LASSO Solver\\[6pt]
{\normalsize A numerical check of the CLA--LASSO identity}}
\author{Thomas Schmelzer}
\date{26 September 2026}

\begin{document}

\maketitle

\begin{abstract}
Schmelzer and Hastie prove that Markowitz's Critical Line Algorithm and the LASSO path
trace the same curve. This companion note checks that result numerically with a stock
LASSO solver, \texttt{lars\_path} from \texttt{scikit-learn}, and no portfolio software.
It covers long--short portfolios under a leverage cap, fixed leverage with varying risk
appetite, and the classical long-only, fully invested frontier. Every corner the path
returns meets its own optimality conditions to about $10^{-13}$ and agrees with
independent quadratic-programming solvers, on factor models of up to 500 assets, on the
Fama--French size and book-to-market portfolios, and on daily returns of 494
S\&P~500 stocks. Called naively, the long-only path stops at the maximum-Sharpe
portfolio. One shift of the response, by an amount computed in advance, lets a single
call reach the minimum-variance portfolio. The script, its checks, the Fama--French
returns and an R cross-check are published, with a locked environment, so that anyone
can repeat them.
\end{abstract}

\section{Introduction}

As the trade-off between expected return and variance moves, the optimal portfolio moves
along straight lines and changes direction only at finitely many \emph{corner
portfolios}, where an asset enters or leaves. Markowitz's Critical Line Algorithm (CLA)
\cite{markowitz1956} walks from one corner to the next. The LASSO \cite{tibshirani1996}
behaves the same way as its penalty moves, and the LARS algorithm \cite{efron2004} walks
its corners, which statisticians call knots. Schmelzer and Hastie \cite{perspective}
prove that after a change of variables the two paths coincide. This note checks that
result numerically, by computing frontiers with a LASSO solver and testing them against
independent solvers, in three settings:
\begin{enumerate}\setlength{\itemsep}{0pt}
\item long--short under a gross-exposure cap, where the plain LASSO path is the frontier
(Section~\ref{sec:longshort});
\item fixed leverage with varying risk appetite, where the same path is rescaled
(Section~\ref{sec:leverage});
\item long only and fully invested, where the nonnegative LASSO path, shifted once,
covers the whole frontier in one call (Section~\ref{sec:longonly}, Algorithm~1).
\end{enumerate}
The usual alternative is to solve a quadratic program on a grid of tilts. Each solve is
only as accurate as its tolerance, and every corner between two grid points is missed.
A path algorithm returns the corners themselves, and every other frontier portfolio is a
linear interpolation between two of them.

\section{Background}
\label{sec:history}

Markowitz \cite{markowitz1952} called a portfolio efficient if no other has a higher
expected return $E$ without a higher variance $V$, or a lower $V$ without a lower $E$. Traced by minimising
$\tfrac12V-\lambda E$ over fully invested portfolios for $\lambda\ge0$, they form the
efficient frontier, with the minimum-variance portfolio at $\lambda=0$. The CLA followed
in 1956 \cite{markowitz1956}. Tobin \cite{tobin1958} and Sharpe \cite{sharpe1964} moved
the picture to the $(\sigma,E)$ plane, where Sharpe located the dominant portfolio, the
one with the highest Sharpe ratio, at the tangent from the riskless rate.

The LASSO fits least squares under $\|\beta\|_1\le t$ and so shrinks and selects at once.
Its path is piecewise linear in $t$. Osborne, Presnell and Turlach \cite{osborne2000}
gave a homotopy for it and Efron et al.\ \cite{efron2004} gave LARS, which also covers
the positive LASSO. Later practice moved to coordinate descent on a grid of penalties
\cite{friedman2010}, which, like a grid of tilts, returns points rather than corners.

The two literatures developed for decades without citing each other
\cite[\S5]{perspective}. The best-known
bridge, Brodie et al.\ \cite{brodie2009}, read a Markowitz problem with a gross-exposure
penalty as a LASSO, but swept the penalty at a fixed target return rather than tracing
the frontier. Fan, Zhang and Yu \cite{fan2012} ran LARS along the gross-exposure cap of
the minimum-variance portfolio. They removed the budget by regressing on a reference
asset, which leaves an $\ell_1$ constraint that is no longer the LASSO's, so their path
is an approximation.

\section{The test problem}
\label{sec:test}

We use the seeded 20-asset example of Figure~1 in \cite{perspective}: a five-factor
covariance $\Sig=\operatorname{diag}(d)+U\operatorname{diag}(\delta)U\trans$, and $\mu$
the sample mean of 50 days of returns simulated from it. Its units are arbitrary and
its expected returns large against its risk, so Sharpe ratios reach $2.1$ per period, far
above anything real assets offer. It serves to illustrate; Section~\ref{sec:evidence}
adds real returns. The companion script
\texttt{python/make\_figures.py} regenerates every figure and number in the note, and its tests
check each against a tolerance. The statement on
\hyperref[sec:availability]{code and data availability} at the end says where to find it.

\section{From mean--variance to least squares}
\label{sec:sub}

Let $\Sig\succ0$ be the covariance of $n$ assets and $\mu$ their expected returns. With
the Cholesky factor $\Sig=LL\trans$, set $X=L\trans$ and $y=L^{-1}\mu$, so that
$X\trans X=\Sig$ and $X\trans y=\mu$. Then for every portfolio $w$
\begin{equation}
\label{eq:square}
\tfrac12\,w\trans\Sig w-\mu\trans w
\;=\; \tfrac12\|y-Xw\|_2^2-\tfrac12\|y\|_2^2 ,
\end{equation}
and under any constraints the mean--variance problem is a least-squares problem.

\begin{lstlisting}
X = np.linalg.cholesky(Sigma, upper=True)  # X.T @ X == Sigma
y = np.linalg.lstsq(X.T, mu)[0]            # X.T @ y == mu
w = np.linalg.lstsq(X, y)[0]               # unconstrained optimum Sigma^{-1} mu
\end{lstlisting}

\noindent
The last line solves the normal equations $X\trans Xw=X\trans y$, which are $\Sig w=\mu$.
With constraints, the optimality conditions become normal equations restricted to the
assets held, a linear system in the trade-off parameter. The weights stay linear until
the holdings change, and those changes are the corners. This is a change of variables,
not a statistical model: $X$ has one row per asset, and nobody measured $y$
\cite[\S4]{perspective}.

Only $X\trans X$ and $X\trans y$ matter, so any square root of $\Sig$ will do. If $\Sig$
is a sample covariance from a $T\times n$ return matrix $R$ with $T>n$, the centred
returns $X=(R-\mathbf 1\bar r\trans)/\sqrt{T-1}$ work, with $y$ from \texttt{lstsq}. For a
factor model, stacking $\operatorname{diag}(d)^{1/2}$ over
$\operatorname{diag}(\delta)^{1/2}U\trans$ gives an $(n+k)\times n$ matrix with no
factorisation at all, and $y=(\operatorname{diag}(d)^{-1/2}\mu,\,0)$. Both reproduce the
Cholesky path to rounding.

\section{Long--short with a leverage cap}
\label{sec:longshort}

\begin{figure}[t]
\centering
\includegraphics[width=\textwidth]{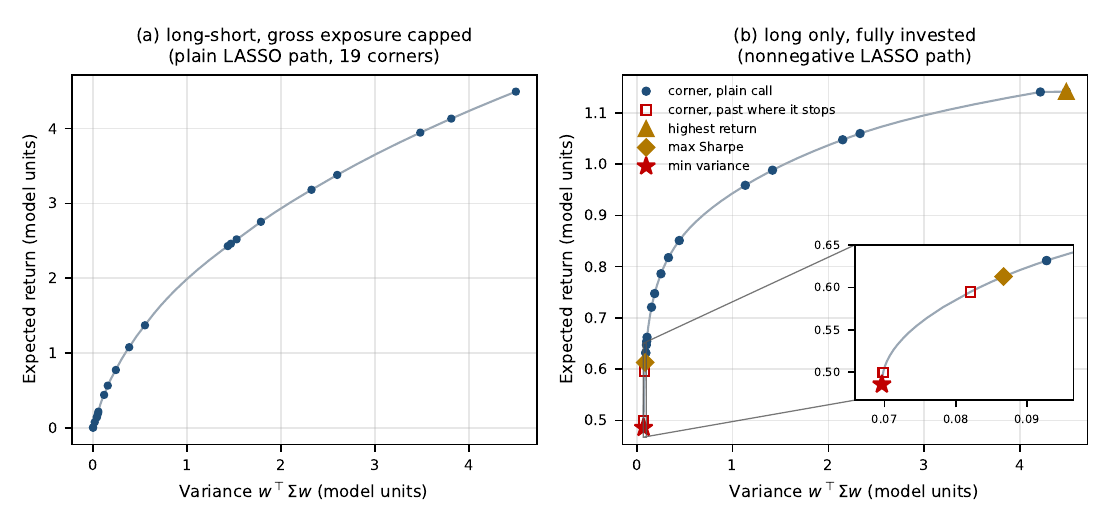}
\caption{Both frontiers of the test problem, computed with a LASSO solver.
\emph{(a)} Long--short under a gross-exposure cap. \emph{(b)} Long-only, fully invested:
blue corners come from a plain call, red squares are the two it misses
(Section~\ref{sec:longonly}). Marked are the highest-return (triangle), maximum-Sharpe
(diamond) and minimum-variance (star) portfolios; the inset magnifies the low-risk end.}
\label{fig:frontier}
\end{figure}

\medskip\noindent
\fbox{\begin{minipage}{\dimexpr\linewidth-2\fboxsep-2\fboxrule\relax}
\textbf{Parameters.} $\lambda$: the tilt, or risk appetite. $c$: the gross-exposure cap,
and $\kappa$: its multiplier, the LASSO penalty of Sections~\ref{sec:longshort}
and~\ref{sec:leverage}. $t=\mathbf 1\trans v=1/\lambda$: the budget of the rescaled
long-only problem, and $\nu$: its multiplier, the LASSO penalty of
Section~\ref{sec:longonly}.
\end{minipage}}
\medskip

With the gross exposure $\|w\|_1$ capped at $c$,
\begin{equation}
\label{eq:mark}
\min_w\ \tfrac12\,w\trans\Sig w-\mu\trans w
\qquad\text{s.t.}\qquad \|w\|_1\le c
\end{equation}
is, by \eqref{eq:square}, the LASSO in constrained form, the observation behind the sparse
portfolios of Brodie et al.\ \cite{brodie2009}. Its penalised form is
$\min_\beta\tfrac12\|y-X\beta\|_2^2+\kappa\|\beta\|_1$, where $\kappa\ge0$ is the
multiplier of the cap and $c=\|\beta(\kappa)\|_1$. So the LASSO path is the frontier of
\eqref{eq:mark} (Theorem~1 of \cite{perspective}), and \texttt{scikit-learn}
\cite{sklearn} computes it:

\begin{lstlisting}
_, _, B = lars_path(X, y, method="lasso")  # columns: start, corners, end
_, _, B = lars_path_gram(Xy=mu, Gram=Sigma, n_samples=len(mu), method="lasso")
\end{lstlisting}

\noindent
The second line gives the same path from $(\Sig,\mu)$ alone. In R the \texttt{lars}
package \cite{lars} traces it too:

\begin{lstlisting}[language=R]
library(lars)
X <- chol(Sigma)                          # t(X) %*% X == Sigma
y <- backsolve(X, mu, transpose = TRUE)   # t(X) %*% y == mu
fit <- lars(X, y, type = "lasso", normalize = FALSE, intercept = FALSE)
B <- t(coef(fit))                         # columns: start, corners, end
\end{lstlisting}

\noindent
It has no positivity constraint, so the long-only path of Section~\ref{sec:longonly}
needs \texttt{scikit-learn}. On the test problem the path has
21 columns: the empty portfolio at the start, 19 corners, each adding one asset, and the
end. The path starts at $w=0$, which is optimal while $\kappa\ge\max_j|\mu_j|$, and ends at
$\kappa=0$, where the cap stops binding at $w^\star=\Sig^{-1}\mu$ with
$\|w^\star\|_1=7.765$. That end maximises $\mu\trans w-\frac12w\trans\Sig w$, the
second-order approximation to expected log growth, so it is the Kelly portfolio
\cite{kelly1956}. Since $\Sig^{-1}\mu$ also maximises the scale-free Sharpe ratio, it
points towards the maximum-Sharpe portfolio. On the test problem the Sharpe ratio rises
from corner to corner, from $0.539$ at the first to $\sqrt{\mu\trans\Sig^{-1}\mu}=2.120$ at
the end; we have not shown that it must. Neither end is the minimum-variance
portfolio, which needs the budget $\mathbf 1\trans w=1$ that \eqref{eq:mark} lacks.

\section{Fixed leverage, varying risk appetite}
\label{sec:leverage}

A mandate usually fixes the leverage, say $c=2$, and varies risk appetite through a tilt
$\lambda>0$. Substituting $w=\lambda v$ in
$\min\tfrac12w\trans\Sig w-\lambda\mu\trans w$ subject to $\|w\|_1\le c$ gives
\eqref{eq:mark} with cap $c/\lambda$. The same path therefore serves: a column $\beta$
with $t=\|\beta\|_1$, scaled to $w=(c/t)\beta$, is optimal at tilt $\lambda=c/t$.

\begin{lstlisting}
t = np.abs(B[:, 1:]).sum(0)             # gross exposure of each column
W, tilts = B[:, 1:] * (c / t), c / t    # the same portfolios at leverage c
\end{lstlisting}

\noindent
At $c=2$ the columns cover tilts from $745.6$ down to $0.2576$. Below that the cap is
slack and the solution $\lambda\Sig^{-1}\mu$ shrinks to zero, so nothing is missing.

\section{Long only, fully invested}
\label{sec:longonly}

\begin{figure}[t]
\centering
\includegraphics[width=0.49\textwidth]{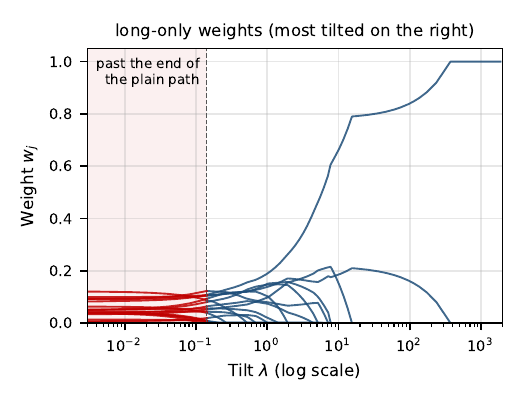}\hfill
\includegraphics[width=0.49\textwidth]{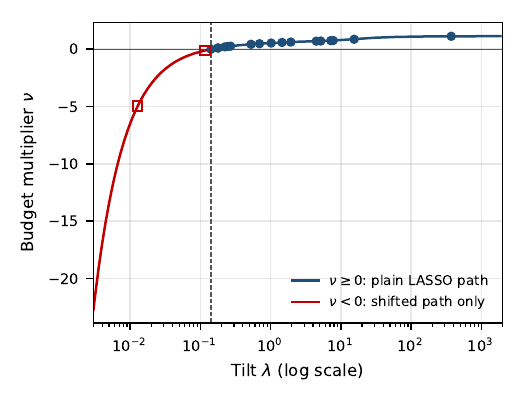}
\caption{The long-only path of the test problem against the tilt $\lambda$. \emph{Left:}
the weights; blue where a plain call supplies them, red (shaded) where it does not.
\emph{Right:} the budget multiplier $\nu=(\mu-\Sig v)_j$, $j$ on the support. A plain
call covers $\nu\ge0$ and ends at $\nu=0$, $\lambda=0.1414$ (dashed); the shift
\eqref{eq:shift} reaches the two further corners (squares) where $\nu<0$.}
\label{fig:weights-lo}
\label{fig:multiplier}
\end{figure}

The classical frontier requires $w\ge0$ and $\mathbf 1\trans w=1$. Substituting
$w=\lambda v$ and using \eqref{eq:square} turns the tilted problem into
\begin{equation}
\label{eq:nnls}
\min_v\ \tfrac12\|Xv-y\|_2^2
\qquad\text{s.t.}\qquad \mathbf 1\trans v=t,\ \ v\ge0,
\qquad t=1/\lambda ,
\end{equation}
with $w=v/t$ (Proposition~3 of \cite{perspective}). For $v\ge0$ the budget is $\|v\|_1$,
so \eqref{eq:nnls} is the nonnegative LASSO, whose penalty $\nu$ is the multiplier of the
budget. \texttt{lars\_path(\dots, positive=True)} traces it, and each column divided by
its sum is a corner portfolio at tilt $1/\mathbf 1\trans v$.

\paragraph{The stop.} Called on $(X,y)$ the path runs from $\nu=\max_j\mu_j$ down to
$\nu=0$ and stops there, because a LASSO solver takes no negative penalty. (If every
$\mu_j\le0$ it returns only the origin.) On the test problem that gives 17 corners, down
to tilt $0.1415$. The stopping point
$v_0=\arg\min_{v\ge0}\frac12v\trans\Sig v-\mu\trans v$ is the end of
Section~\ref{sec:longshort} with long-only weights. Writing $v=sw$ with
$\mathbf 1\trans w=1$ and maximising over the scale $s\ge0$ first leaves
$\frac12(\mu\trans w)^2/w\trans\Sig w$, half the squared Sharpe ratio, so
$v_0/\mathbf 1\trans v_0$ is the long-only maximum-Sharpe portfolio, at tilt $0.1414$.
The frontier goes on, though: the budget still binds while $\nu$ turns negative
(Figure~\ref{fig:multiplier}), which a constraint allows and a penalty does not
\cite[\S3.4]{perspective}. The stop comes from the penalised form that LASSO solvers use, not from the problem.

\paragraph{The shift.} Completing the square once more moves the penalty into the
response:
\begin{equation}
\label{eq:shift}
\tfrac12\|Xv-y\|_2^2+\nu\,\mathbf 1\trans v
=\tfrac12\bigl\|Xv-(y-\nu d)\bigr\|_2^2+\text{const},
\qquad X\trans d=\mathbf 1 .
\end{equation}
If the nonnegative LASSO runs on $y-\nu_0d$ for some $\nu_0<0$, its penalty $\nu'\ge0$
acts as $\nu=\nu_0+\nu'$ on the original problem, and one path covers every $\nu\ge\nu_0$.
For the Cholesky factor $d=(X\trans)^{-1}\mathbf 1$; for a tall $X$ least squares returns
the minimum-norm solution. \texttt{lars\_path} reports $\nu'$ divided by the number of
rows, and it stops after 500 steps unless told otherwise, which a large problem needs:

\begin{lstlisting}
d = np.linalg.lstsq(X.T, np.ones(len(mu)))[0]   # solves X.T @ d == 1
alphas, _, V = lars_path(X, y - nu0 * d, method="lasso", positive=True,
                         max_iter=10 * len(mu))
nu = nu0 + len(y) * alphas                      # budget multiplier per column
corners = V[:, 1:-1]                            # drop the origin and the end
W, tilts = corners / corners.sum(0), 1 / corners.sum(0)
\end{lstlisting}

\paragraph{Choosing $\nu_0$.} Any $\nu_0$ below the last corner works, and the last
corner is known in advance. Beyond it the portfolio holds the minimum-variance support
$S$, which is the support of $\arg\min_{u\ge0}\|Xu-d\|_2$ \cite{lawson1974}. There
$v_S=a-\nu b$ with $a=\Sig_{SS}^{-1}\mu_S$ and $b=\Sig_{SS}^{-1}\mathbf 1$, and the
segment stays optimal while $\nu<a_i/b_i$ for $b_i>0$ and, for $j\notin S$ with
$\Sig_{jS}b>1$, while $\nu\le(\mu_j-\Sig_{jS}a)/(1-\Sig_{jS}b)$. The smallest bound is the
last corner $\nu_{\mathrm{last}}$:

\begin{lstlisting}
S = nnls(X, d)[0] > 1e-9                         # minimum-variance support
a = np.linalg.solve(Sigma[np.ix_(S, S)], mu[S])
b = np.linalg.solve(Sigma[np.ix_(S, S)], np.ones(S.sum()))
h = 1 - Sigma[np.ix_(~S, S)] @ b
g = mu[~S] - Sigma[np.ix_(~S, S)] @ a
nu_last = min((a[b > 0] / b[b > 0]).min(),
              (g[h < 0] / h[h < 0]).min(initial=np.inf))
nu0 = nu_last - max(abs(nu_last), np.abs(mu).max())
\end{lstlisting}

\noindent
No fixed value would be safe. $\nu_{\mathrm{last}}$ is $-4.95$ on the test problem,
$-11.9$ on one of the models of Table~\ref{tab:robust}, $-0.178$ on the stock data and
$-0.0118$ on the portfolio data of Section~\ref{sec:evidence}, where $\mu$ is small. The margin $\delta$ in $\nu_0=\nu_{\mathrm{last}}-\delta
\max(|\nu_{\mathrm{last}}|,\|\mu\|_\infty)$ has to be large enough that the end of the
path does not merge with the last corner, and small enough not to cost accuracy, which
falls roughly in proportion to $|\nu_0|$. On the stock data $\delta=10^{-4}$ loses the
last corner; from $\delta=10^{-2}$ to $1$ all 94 are found with KKT residuals below
$4\times10^{-13}$, and at $\delta=100$ the residual has grown to $2\times10^{-11}$. On the
portfolio data even $\delta=10^{-2}$ loses the last corner, and $\delta=1$ finds all 12. The
code takes $\delta=1$. The minimum-variance portfolio itself is the limit $\nu\to-\infty$; on the
final segment it is $w_S\propto\Sig_{SS}^{-1}\mathbf 1$.

\paragraph{What $\nu$ means.} By \eqref{eq:shift} the point with multiplier $\nu$ solves
the same problem with $\mu$ replaced by $\mu-\nu\mathbf 1$, so it is the long-only
maximum-Sharpe portfolio at risk-free rate $r=\nu$, Black's zero-beta rate
\cite{black1972}. That is classical. The LASSO view adds that the rate is the path's own
parameter, so for $\nu_0<r<\max_j\mu_j$ the maximum-Sharpe portfolio is an interpolation
between the two columns that bracket $r$:

\begin{lstlisting}
assert nu[-1] < r < nu[0]                        # the path covers the rate r
k = np.argmax(nu < r)                            # first column past r
s = (nu[k - 1] - r) / (nu[k - 1] - nu[k])
w_r = V[:, k - 1] + s * (V[:, k] - V[:, k - 1])
w_r /= w_r.sum()                                 # max-Sharpe portfolio at r
\end{lstlisting}

\medskip\noindent
\fbox{\begin{minipage}{\dimexpr\linewidth-2\fboxsep-2\fboxrule\relax}
\textbf{Weights versus frontier.} Between corners the weights are linear in the tilt
$\lambda$ (or, in Section~\ref{sec:longshort}, in the cap $c$) and the variance is
quadratic, so each piece of the frontier is an arc.
Traced by the tilt $\lambda$ at fixed constraints (Sections~\ref{sec:leverage}
and~\ref{sec:longonly}) it has no kinks either, since its slope
$\mathrm dE/\mathrm dV=1/(2\lambda)$ varies continuously. The corners then show only in
the weights (Figure~\ref{fig:weights-lo}). The long--short path of
Section~\ref{sec:longshort} is traced by the cap instead and does kink
(Figure~\ref{fig:frontier}(a)).
\end{minipage}}
\medskip

\medskip\noindent
\fbox{\begin{minipage}{\dimexpr\linewidth-2\fboxsep-2\fboxrule\relax}
\textbf{Algorithm 1: the long-only frontier from \texttt{lars\_path}.}
\begin{enumerate}\setlength{\itemsep}{0pt}
\item Build $X,y$ with $X\trans X=\Sig$, $X\trans y=\mu$, or use \texttt{lars\_path\_gram}
with \texttt{Xy=mu - nu0}.
\item Solve $X\trans d=\mathbf 1$, compute $\nu_{\mathrm{last}}$ and set $\nu_0$ as above.
\item Run the positive LASSO on $y-\nu_0d$ with \texttt{max\_iter} well above $n$, and
recover $\nu=\nu_0+n_{\mathrm{rows}}\alpha$.
\item Drop the first and last columns; each remaining $v/\mathbf 1\trans v$ is a corner,
at tilt $1/\mathbf 1\trans v$.
\item Take the minimum-variance end as $w_S\propto\Sig_{SS}^{-1}\mathbf 1$, and any other
point by interpolation, in $1/\lambda$ for a tilt or in $\nu$ for a rate.
\end{enumerate}
\end{minipage}}
\medskip

\section{Numerical evidence}
\label{sec:evidence}

Two kinds of evidence are reported. The first needs no second solver: each portfolio
the path returns is checked against its own optimality conditions. For the cap with
multiplier $\kappa$ these are $(\mu-\Sig w)_j=\kappa\,\mathrm{sign}(w_j)$ on the support
and $|(\mu-\Sig w)_j|\le\kappa$ off it; for the long-only problem at tilt $\lambda$ they are
$(\Sig w-\lambda\mu)_j=\gamma$ on the support and $\ge\gamma$ off it, with
$\mathbf 1\trans w=1$ and $w\ge0$. The KKT residual is the worst violation, divided by
$\|\mu\|_\infty$ or by the size of the gradient. The second compares with baselines that
are independent of the path. The long--short problems go to Clarabel, an interior-point
solver, through \texttt{cvxpy}; the long-only problem goes to \texttt{quadprog}, the
active-set method of Goldfarb and Idnani \cite{goldfarb1983}.

\emph{Long--short.} Every column has a KKT residual below $2.6\times10^{-15}$, and
below $2.9\times10^{-15}$ once rescaled to fixed leverage. At each column's cap Clarabel matches the path to
$3.0\times10^{-7}$ at the corners and $1.4\times10^{-11}$ inside segments. The gap at the
corners is the baseline's, since interior-point methods are least accurate where the
holdings change. At fixed leverage the figures are $6.2\times10^{-6}$ and
$1.9\times10^{-11}$.

\emph{Long only.} The shifted path finds 19 corners on the test problem, the 17 of a
plain call and two more at tilts $0.1158$ and $0.0128$. They match the QP to
$6.0\times10^{-13}$, points inside segments to $7.9\times10^{-14}$, the
minimum-variance end to $1.0\times10^{-16}$, and the maximum-Sharpe portfolio at $r=0$
to $3.4\times10^{-15}$. The corners and the minimum-variance end have KKT residuals below
$3.6\times10^{-14}$. Table~\ref{tab:robust} repeats both checks on other sizes and seeds.
Where the two disagree, at $n=500$, the residual shows that the gap of
$6.7\times10^{-11}$ is the QP's.
The number of corners need not equal the number of assets, because some assets leave
and re-enter.

\begin{table}[t]
\centering
\small\setlength{\tabcolsep}{4.5pt}
\begin{tabular}{rccccccr}
\toprule
$n$ & seeds & corners & $\nu_{\mathrm{last}}$ & KKT residual & worst corner
& min.-variance end & time (ms)\\
\midrule
20 & 3 & 19 & $-0.57$ to $0.12$ & $1.1\times10^{-14}$ & $8.6\times10^{-14}$
& $2.4\times10^{-16}$ & $0.8$\\
50 & 3 & 49 & $-11.9$ to $-0.74$ & $1.0\times10^{-13}$ & $1.5\times10^{-13}$
& $1.7\times10^{-16}$ & $2.5$\\
100 & 3 & 99--103 & $-0.56$ to $-0.38$ & $2.5\times10^{-14}$ & $1.1\times10^{-11}$
& $3.3\times10^{-16}$ & $7.7$\\
500 & 3 & 503 & $-0.45$ to $-0.26$ & $2.1\times10^{-13}$ & $6.7\times10^{-11}$
& $4.3\times10^{-16}$ & $239$\\
\bottomrule
\end{tabular}
\caption{Algorithm~1 on five-factor models, three seeds each: the worst KKT residual over
the corners and the minimum-variance end, the largest $\ell_\infty$ weight discrepancy
against an independent active-set QP at the same tilt, and the median time of one run of
Algorithm~1, $\nu_{\mathrm{last}}$ included, on a laptop.}
\label{tab:robust}
\end{table}

\emph{Real returns, reproducible.} The daily value-weighted returns of the 100
portfolios that Fama and French form on size and book-to-market \cite{french}, from
January 2016 to December 2025, are published with the code. One portfolio misses 250
days and is dropped, which leaves 99 portfolios over 2514 days. With the centred returns
as $X$ the last corner sits at $\nu_{\mathrm{last}}=-0.0118$, eleven times
$\max_j|\mu_j|$ below zero. Algorithm~1 returns all 12 corners. They match the QP to
$2.1\times10^{-13}$ and have KKT residuals below $3.9\times10^{-14}$, and the
minimum-variance portfolio, which holds 7 of the portfolios, matches to
$1.6\times10^{-15}$. Diversified portfolios are highly correlated, so the frontier turns
only 12 times and a grid of QPs is cheap here. The stock data test the method at scale.

\emph{Real returns at scale.} On daily returns of 494 S\&P~500 stocks over 1213 days (July 2021
to May 2026, from Yahoo Finance; the script that fetches them is published with the code), with the centred returns as $X$, the last corner sits at
$\nu_{\mathrm{last}}=-0.178$, about fifty times $\max_j|\mu_j|$ below zero. Algorithm~1
returns all 94 corners in $0.14$ seconds, while one QP takes about $0.12$ seconds, so a
100-point grid costs about 12 seconds and still misses the corners. Warm-starting
helps less than one might hope: OSQP \cite{stellato2020}, set up once so that its
factorisation is reused, with each tilt changing only the linear term and each solve
starting from the last, takes 4.9 seconds for the same grid at tolerance $10^{-6}$ and
agrees with the cold QP to $4.1\times10^{-13}$ at every grid point. At $10^{-5}$ it takes
3.4 seconds but misses some points by $10^{-3}$. Every corner has a KKT
residual below $3.7\times10^{-13}$ and matches its QP to $5.4\times10^{-13}$, and the
minimum-variance portfolio, which holds 53 stocks, matches to $1.2\times10^{-15}$. The timings come from a laptop.

\section{When to use it}
\label{sec:scope}

The path beats a grid of QP solves, cold- or warm-started, on both speed and accuracy. LARS is in R
and \texttt{scikit-learn}, the positive LASSO only in \texttt{scikit-learn}, so corner
portfolios need no portfolio software, and one call
shows that the knots of a LASSO path and the turning points of a frontier are the same
points. It is the wrong tool when $\Sig$ is singular,
where the identity is not known to survive \cite[\S7]{perspective}, or when the portfolio
needs more than one linear constraint. The LASSO that \texttt{scikit-learn} solves has
exactly one besides the signs, $\|\beta\|_1\le t$: the gross-exposure cap of
Sections~\ref{sec:longshort} and~\ref{sec:leverage}, or, with \texttt{positive=True},
the budget $\mathbf 1\trans w=1$ of Section~\ref{sec:longonly}. The signs are either all
free or all nonnegative. A dollar-neutral book adds the equality $\mathbf 1\trans\beta=0$,
and a $130/30$ book needs the budget and the gross cap at once. Path algorithms for a
LASSO with general linear constraints exist \cite{gaines2018,james2020}, but not in
\texttt{scikit-learn}. A risk budget $w\trans\Sig w\le\sigma^2$ has no path of its own, but
along each segment the variance is quadratic in the parameter, so the portfolio at a
given $\sigma$ comes from one scalar equation.

The numbers above are evidence for Theorem~1 and Proposition~3 of \cite{perspective},
not a proof of them; the proofs are there. They test what each result predicts.
Theorem~1 says the LASSO path is the frontier under a gross-exposure cap: every column
meets the optimality conditions of its cap and matches Clarabel there. Proposition~3
says the nonnegative LASSO is the long-only frontier: every corner, and every point
between corners, matches an active-set QP at the same tilt. Both make the frontier
piecewise linear in its parameter, so a path finds its corners exactly where a grid of QP
solves, cold- or warm-started, misses them. The Fama--French returns are committed with
the code, so that real-data check reruns on every build. On the stock data the checks pass on the
note's snapshot and on a later download, although the two differ in the number of
corners and in $\nu_{\mathrm{last}}$. The checks were written by an author of
\cite{perspective}, so they are not independent in the usual sense. The baseline solvers
are, and the script and its tolerances are published so that anyone can rerun them.
Along the way the note adds what a practitioner needs to use the result: the shift, a
$\nu_0$ computed in advance, and a \texttt{max\_iter} large enough for big problems.

\section*{Code and data availability}
\label{sec:availability}

The software is at \url{https://github.com/Jebel-Quant/lasso}, in the directory
\texttt{preprints/lasso/}, under the MIT license. \texttt{python/make\_figures.py} regenerates every
figure and number in the note; \texttt{uv run python/make\_figures.py test} runs its checks. It
is a single Python script that declares its own pinned dependencies, and
\texttt{python/make\_figures.py.lock} next to it fixes Python 3.12 and every package it
pulls in, to the exact version and hash. The R snippet of
Section~\ref{sec:longshort} is \texttt{R/long\_short.R}, with a check against
\texttt{lars\_path}.

The lockfile fixes the software, not the hardware. Residuals and gaps near $10^{-13}$ come
out with different last digits on another CPU or BLAS build, and the timings scale with
the machine, so a rerun will not match the printed values digit for digit. The tests
therefore check the residuals and gaps against tolerances between $10^{-12}$ and
$10^{-9}$, and leave the timings unchecked.

The test problem and the factor models of Table~\ref{tab:robust} are simulated inside
the script from fixed seeds, so they need no data. The portfolio returns of
Section~\ref{sec:evidence} are in the repository as \texttt{data/french100\_daily.csv},
cut from Kenneth French's Data Library \cite{french} (CRSP vintage of August 2026,
SHA-256 \texttt{98298ef6\ldots3c43}), and \texttt{python/fetch\_french.py} rebuilds them. The
library rebuilds its history each month from CRSP, so a later download can differ where
CRSP revises a return. These data remain Fama and French's and are not covered by the
repository's MIT license. The stock returns of
Section~\ref{sec:evidence} come from Yahoo Finance and are not redistributed.
\texttt{python/fetch\_sp500.py} downloads them: the S\&P~500 constituents listed on Wikipedia,
adjusted closing prices from 2021-06-01 to 2026-06-01, keeping names with no more than
5\% of days missing. The note used the snapshot with SHA-256 \texttt{b5faa522\ldots9937e}.
A later download differs slightly, because the constituents change and Yahoo revises
adjusted prices; the checks pass on it all the same.

\section*{Acknowledgements}

I thank Trevor Hastie for many helpful discussions.

\end{document}